\documentclass{Interspeech2024}

\interspeechcameraready

\title{Iterative Prototype Refinement for Ambiguous Speech Emotion Recognition}

\name[affiliation={1}]{Haoqin}{Sun}
\name[affiliation={1}]{Shiwan}{Zhao}
\name[affiliation={2}]{Xiangyu}{Kong}
\name[affiliation={1}]{Xuechen}{Wang}
\name[affiliation={1}]{Hui}{Wang}
\name[affiliation={1}]{Jiaming}{Zhou}
\name[affiliation={1,*}]{\\Yong}{Qin}

\address{
  $^1$Nankai University, Tianjin, China\\
  $^2$University of Leicester, Leicester, United Kingdom}
\email{qinyong@nankai.edu.cn}


\begin{document}

\maketitle
\renewcommand{\thefootnote}{}
\footnotetext{* Corresponding author.}

\begin{abstract}
    Recognizing emotions from speech is a daunting task due to the subtlety and ambiguity of expressions. Traditional speech emotion recognition (SER) systems, which typically rely on a singular, precise emotion label, struggle with this complexity. Therefore, modeling the inherent ambiguity of emotions is an urgent problem. In this paper, we propose an iterative prototype refinement framework (IPR) for ambiguous SER. IPR comprises two interlinked components: contrastive learning and class prototypes. The former provides an efficient way to obtain high-quality representations of ambiguous samples. The latter are dynamically updated based on ambiguous labels—the similarity of the ambiguous data to all prototypes. These refined embeddings yield precise pseudo labels, thus reinforcing representation quality. Experimental evaluations conducted on the IEMOCAP dataset validate the superior performance of IPR over state-of-the-art methods, thus proving the effectiveness of our proposed method.
\end{abstract}

\section{Introduction}

The advancement of affective computing \cite{hu2018touch, arsikere2014computationally} has spurred the development of a wide array of emotion recognition corpora \cite{haq2009speaker, busso2016msp, livingstone2018ryerson}. Despite this progress, the intricate and costly process of data annotation, combined with the inherent ambiguity of emotional expressions, poses significant challenges. Traditional speech emotion recognition (SER) systems commonly rely on the assumption that emotions can be clearly categorized, attributing a precise or singular label to each vocal expression. For instance, Sun et al. \cite{sun2023fine} have engaged in representation learning with IEMOCAP \cite{busso2008iemocap}, while Liu et al. \cite{liu2023discriminative} have explored metric learning on EmoDB \cite{burkhardt2005database} and CAISA \cite{zhang2008design}. Nevertheless, these approaches may fall short of capturing the full spectrum of emotional ambiguity. This observation is supported by psychological and statistical research \cite{ortony2022cognitive, tao2009multiple}, as well as machine learning studies \cite{wang23q_interspeech}, which collectively suggest that the boundaries between different emotional categories are not always distinct. The complexity and nuance of emotional expressions, coupled with the diversity in corpus composition, demand more sophisticated SER methodologies capable of more effectively navigating the intricate emotional landscape. Such advancements would transcend the constraints of singular emotion labeling, thereby enhancing the precision and broad applicability of emotion recognition across a varied range of emotional states.

Recently, researchers have attempted to address the challenges in ambiguous SER by modeling emotion ambiguity. For example, Lotfian et al. \cite{lotfian2018predicting} propose a multitask learning framework to learn the primary and secondary emotions of an utterance, ignoring information about other minor emotions. In contrast, Ando et al. \cite{ando2018soft} propose soft-target label learning -- estimating the proportion of all emotions, which increases the complexity of soft label learning. Moreover, Fei et al. \cite{fei2020topic} propose a multi-label emotion classification method to represent the ambiguity of emotions. However, these methods are primarily applicable to corpora retaining expert voting information during annotation. In the absence of annotated records, the generalizability of these methods is challenged. 

To address these challenges, Zhou et al. \cite{zhou2022multi} propose a Multi-Classifier Interaction Learning (MCIL) framework. This framework emulates the expert annotation process through multiple classifiers. On the one hand, it represents the annotation process using the classification results of multiple classifiers, facilitating soft label learning. On the other hand, it employs the majority voting mechanism to generate a precise label, enabling the study of traditional SER techniques. It is important to note that the efficacy of MCIL is heavily contingent on the performance of individual classifiers. Poor performance of these classifiers may result in unsatisfactory data labeling quality and emotion recognition performance. Conversely, if a particular classifier exhibits exceptional performance, the effectiveness of interactive learning may be diminished. Furthermore, simply using classifiers to annotate ambiguous samples and retraining the model may compromise the discriminability of the model in this type of data.

Inspired by the work \cite{zhou2023madi}, in this paper, we propose an iterative prototype refinement framework (IPR) for ambiguous SER. IPR comprises three key phases: class prototype learning, class prototype updating, and contrastive learning. In the class prototype learning phase, precise samples facilitate the initial acquisition of class prototype embeddings via a warm-up mechanism. Subsequently, during the class prototype updating phase, unlabeled ambiguous samples are introduced to participate in training process. We calculate the distance of the samples from each class prototype to construct ambiguous soft labels and update the class prototypes by the proportion of classes in the soft labels. Finally, contrastive learning is employed to augment the discriminative capability of the model, updating class prototypes when ambiguous samples and their enhanced counterparts points to one prototype. The main contributions are summarized as follows:

\begin{figure*}[t]
  \centering
  \includegraphics[width=6.5in]{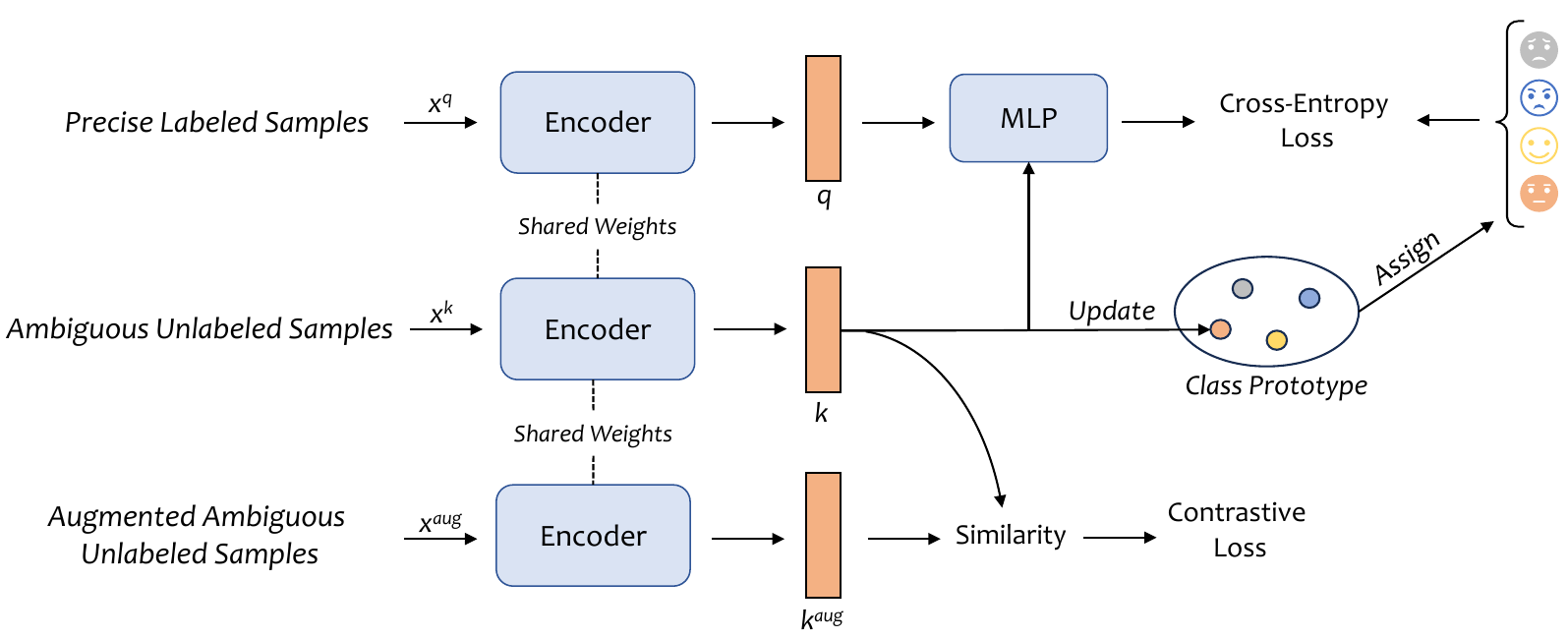}
  \caption{The overall architecture of IPR, which consists of class prototype learning phase, class prototype updating phase and contrastive learning phase.}
  \label{fig:overview}
\end{figure*}

(1) We propose a novel iterative prototype refinement framework for ambiguous SER. Contrastive learning aids in the learning of ideal representations, improving the quality of class prototypes. Subsequently, enhanced class prototypes guide the model in generating high-quality representations.

(2) IPR is trained on a small amount of high-quality data, enabling the construction of ambiguous soft labels and single labels for a large amount of data, thereby possessing better generalization capability.

(3) Experimental results on IEMOCAP demonstrate that IPR outperforms the current state-of-the-art methods, achieving an accuracy of 70.75\%, with an absolute improvement of 2.00\%.

\section{Proposed Methods}
Illustrated in Fig.~\ref{fig:overview}, our proposed framework, IPR, comprises three key phases: class prototype learning, class prototype updating, and contrastive learning.

\subsection{Class Prototype Learning}
First, we initially train the encoder using the precise labeled samples $x^{q}_i$. Meanwhile, for each class $c \in \{1,2,...,C\}$, the prototype vectors $p_c$ are initialized based on the precise samples as representative embeddings.
\begin{align}
q_i=&\operatorname{Encoder}\left(x^{q}_i ; \theta\right),\\
p_{c}=&\frac{1}{n} \sum_{i \in c} q_i,
\end{align}
where $\operatorname{Encoder}$ represents the pre-trained model Wav2vec2.0 \cite{baevski2020wav2vec}, which is commonly used as a feature extractor \cite{dang2022using, gao2024enhancing}. $\theta$ represents the trainable parameters. $n$ represents the number of samples belonging to class $c$.

Since, when the class prototype is not adequately learned, it has a negative impact in guiding the model to learn unlabeled data. Therefore, we set up the warm-up mechanism so that the encoder learns the better representation of the class prototypes on the precise labeled data.

\subsection{Class Prototype Updating}
Once we acquire a prototype representation trained on precise labeled samples, the training process of the model will incorporate ambiguous unlabeled samples $x^{k}_i$. Then, we compute the similarity between $k_i$ and each class prototype vector $p_c$ to construct ambiguous soft label $\{s_1,s_2,...,s_C\}$,  where $s_c$ is ${k}^{\top}_i {p}_{c}$. The class prototype vector with the largest similarity to $k_i$ is the pseudo label $z_c$ of that representation.
\begin{eqnarray}
z_{c} = \arg\max {k}^{\top}_i {p}_{c},
\end{eqnarray}
where $\boldsymbol{p}_{c}$ represents the $c$-th class. $\max$ represents maximization operation.

In the subsequent step, following the assignment of pseudo labels to ambiguous unlabeled samples, we dynamically adjust all class prototype vectors based on the ambiguous labels to incorporate the emotional information introduced by these samples. Traditional methods for calculating prototype embeddings require iterative updates, which is time-consuming and computationally intensive. To avoid this problem, we use moving average as the update method for prototype embedding.
\begin{align}
p_{c}&=\operatorname{Normalize}\left(\gamma p_{c}+(1-\gamma) {k_i}{s_c}\right), \\ \notag
\quad \text { if } c&=\max\operatorname{MLP}\left(k_i\right)=\max\operatorname{MLP}\left(k^{aug}_i\right),
\end{align}
where Normalize represents the normalization operation. $\gamma$ represents a tunable hyperparameter. MLP stands for the classifier. 

\subsection{Contrastive learning}
To enhance the quality of learned emotion representations in ambiguous samples, we integrate contrastive learning into IPR. For each ambiguous sample $x^k_i$, we initiate the process by generating an augmented version, denoted as $x^{aug}_i$, through a mixing method. This method encompasses various techniques such as adding noise, changing volume, adding reverberation, and changing pitch. Subsequently, we employ class prototype embeddings to assign pseudo labels for both $x^k_i$ and $x^{aug}_i$. In a batch of 2$N$ samples, comprising $x^k_i$ and $x^{aug}_i$, we differentiate between positive examples, which share the same pseudo label, and negative examples, which possess different pseudo labels. The loss is computed as follows:
\begin{align}
\mathcal{L}_{con}=-\log \frac{\exp \left(\left({x^k_i} \cdot {x^{aug}_i}\right) / \tau\right)}{\sum_{j \in \{k,aug\}} \exp \left(\left({x^k_i} \cdot {x^{aug}_j}\right) / \tau\right)},
\end{align}
where $\tau$ is a temperature parameter.

\subsection{Training Objective}
To train our proposed IPR, three distinct loss functions are employed: cross-entropy loss for precise labeled samples, cross-entropy loss for ambiguous samples, and contrastive learning loss. The overall training loss function is formulated as:
\begin{align}
\mathcal{L}=\alpha \mathcal{L}_{c l s}\left({x^q},y\right)+\beta \mathcal{L}_{c l s}\left({x^k},\hat{y}\right) + \mu\mathcal{L}_{con},
\end{align}
where $\alpha$, $\beta$, and $\mu$ represent the tunable trade-off factors. $y$ and $\hat{y}$ represent ground-truth and pseudo labels, respectively.

\section{Experiments}
\subsection{Dataset}
IEMOCAP \cite{busso2008iemocap} stands out as a widely used dataset in SER. Developed by the Signal Analysis and Interpretation Laboratory at the University of Southern California, this dataset comprises emotionally charged dialogues enacted by 10 actors representing diverse demographics in terms of gender and age. These dialogues are orchestrated using constraint-based scripts crafted to mimic authentic interaction scenarios. we focus on four emotions: anger, happiness, sadness, and neutral, where excitement class is merged into the happiness class.

\begin{table}[t]
\caption{The sizes of training and testing datasets.}
  \label{tab:dataset}
  \centering
\begin{tabular}{cccc}  \hline
Dataset & D1   & D2   & D3  \\ \hline
IEMOCAP & 1710 & 3421 & 400 \\ \hline
\end{tabular}
    \vspace{-6pt}
\end{table}

\subsection{Experimental Settings}
As shown in Table~\ref{tab:dataset}, following the methodology proposed by \cite{zhou2022multi}, we partition the IEMOCAP dataset into three distinct subsets tailored for addressing ambiguous SER scenarios. Specifically, D1 encompasses precise labeled samples exhibiting minimal ambiguity in emotional classification. D2 comprises unlabeled samples with moderate ambiguity in emotion. D3, on the other hand, includes labeled samples characterized by the highest degree of ambiguity in emotion, serving as the evaluation set. The determination of emotion ambiguity is predicated on the level of consensus among experts during the annotation process. For each utterance in IEMOCAP, multiple experts contribute their assessments, and the consistency of their judgments dictates the attribution of the sample. A high level of agreement among annotators results in D1, whereas increased variability in assessments places a sample within D2. Samples demonstrating substantial disagreement among annotators are designated to D3.

In our experimental setup, we establish three baseline systems: a baseline system, a baseline$^+$ system, and a supervised baseline system. The baseline system is trained exclusively on D1 and subsequently evaluated on D3. In contrast, the baseline$^+$ system utilizes both D1 and unlabeled D2 during training, which employs model-generated soft labels instead of the prototype-assigned pseudo labels employed in our proposed method. Finally, the supervised baseline system is trained on a combination of labeled D1 and D2 data and tested on D3. This system consists of an encoder and a MLP without the incorporation of class prototype and contrastive learning.

Our proposed framework, IPR, is trained on A100 utilizing PyTorch. We adopt a batch size of 8 and set the maximum training epoch to 50, incorporating a warm-up mechanism for the initial 10 epochs. We employ the AdamW optimizer with an initial learning rate of $10^{-5}$, setting $\gamma$ to 0.99. The trade-off parameters $\alpha$ and $\mu$ are fixed at 1.0 and 0.2, respectively. Initially, $\beta$ is set to 0.0 and updated according to Eq. 8, gradually increasing based on the current and maximum epoch numbers. The increment of $\beta$ accelerates with each epoch until reaching 0.5.
\begin{align}
\beta = \min \left(\frac{\text{weight}_{m} \times\left(r^{\frac{ \text{epoch}_c}{\text{epoch}_{m} / 2}-1}\right)}{r-1},\text{weight}_{m}\right),
\end{align}
where the weight parameter weight$_m$ is set to 0.5, denoting its maximum value. Here, epoch$_m$ represents the maximum epoch, while epoch$_c$ denotes the current epoch. The exponential growth factor $r$ governs the rate of increase. Consistent with prior research \cite{zhou2022multi}, we adopt accuracy (Acc) as the primary metric for assessing the efficacy of our proposed method. The final Acc is the average results across five seeds.

\begin{table}[t]
  \caption{Performance comparison of our proposed methods with SOTA approaches on IEMOCAP.}
  \label{tab:sota}
  \centering
  \begin{tabular}{ccc}
    \toprule
     \textbf{Method} & \textbf{Description} & \textbf{Acc(\%)}\\
    \midrule
    Cummins et al.\cite{cummins2017image}& AlexNet & 59.20\\
    Li et al. \cite{li2018attention} & CNN & 58.00\\
    Ando et al. \cite{ando2019speech}&  Multi-label & 57.80\\
    Liu et al. \cite{liu2020speech} & CapsNet & 58.55\\
    Zhou et al. \cite{zhou2022multi} & Majority voting & 65.00\\
    Zhou et al. \cite{zhou2022multi} & MCIL  & 67.00\\
    \midrule
    Ghifary et al. \cite{ghifary2014domain} & DaNN & 65.00\\
    Yu et al. \cite{yu2019transfer} & DAAN & 65.75\\
    Ganin et al. \cite{ganin2015unsupervised}&  DANN & 68.50\\
    Cui et al. \cite{cui2020towards} & BNM & 68.75\\
    \midrule
    \textbf{IPR} & \textbf{Prototype learning} & \textbf{70.75}\\
    \bottomrule
  \end{tabular}
  \vspace{-6pt}
\end{table}

\subsection{Comparison with State-of-the-Art Methods}
To demonstrate the effectiveness of IPR, we compare IPR with many state-of-the-art (SOTA) methods. Table~\ref{tab:sota} lists a brief description and accuracy results of these methods, where the first half is SER method, retrained following MCIL training algorithm, and the second half is the classical semi-supervised methods. 

We could find that our method achieves the optimal performance on Acc, reaching 70.75\%, which is significantly better than SOTA methods (with an absolute improvement of 2.00\%). This performance is mainly attributed to the collaborative effect of class prototype learning and contrastive learning, which reinforce each other, thus allowing the model to capture weak dominant emotions from ambiguous utterances. In comparison with the SER approaches, IPR demonstrates effectiveness and reliability in generating pseudo labels. With the guidance of pseudo labels, IPR achieves a 3.75\% improvement on Acc. Furthermore, in the comparison with the semi-supervised methods, we find that IPR is improved by 2.00\% on Acc. These semi-supervised methods only utilize a large amount of unlabeled data for the purpose of improving recognition performance. It does not seem to tackle the problem of annotation on large amounts of data by learning on a small number of precise samples. The experimental results provide strong evidence that our proposed IPR could provide ideal performance for ambiguous SER and permit more reliable data annotation.

\subsection{Baseline System Analysis}
We report the experimental results of IPR with different baseline systems as shown in Table~\ref{tab:results}. First, we observe that by integrating these unlabeled samples, the performance of IPR improves significantly, by 6.50\% over baseline on Acc. This highlights the importance of using unlabeled ambiguous data to optimize model. Second, IPR outperforms baseline$^+$. This proves that IPR, an iterative prototype refinement framework, outperforms model-generated soft labels in the generation of pseudo labels. Notably, the baseline$^+$ also outperforms the majority voting method of MCIL in SOTA. Thus, when faced with a corpus without voting information, our method utilizes a small amount of data for training and could achieve a large number of data annotation, both single and ambiguous soft labels. Finally, by comparing IPR and supervised baseline, we could notice that on Acc, the proposed method reduces 1.55\%, while MCIL reduces 5.30\%. As a result, IPR is closer to the performance of supervised system, while it produces more precise and reliable pseudo labels when annotating unlabeled samples.

\begin{table}[t]
  \caption{Experimental results of different baseline systems on IEMOCAP dataset. The baseline$^+$ represents training using model-generated pseudo labels. $\star$ indicates that p-value $<$ 0.05 (compared with baseline). $\bigstar$ indicates that p-value $<$ 0.05 (compared with baseline and baseline$^+$).} 
  \label{tab:results}
  \centering
  \begin{tabular}{cccc}
    \toprule
    \textbf{Method} & \textbf{Train}  & \textbf{Acc(\%)}\\
    \midrule
    baseline & D1  & 64.25\\
    baseline$^+$ & D1 and unlabeled D2  & 66.85$^{\star}$\\
    \textbf{IPR} &  \textbf{D1 and unlabeled D2}  &  \textbf{70.75}$^{\bigstar}$\\
    supervised baseline & both labeled D1 and D2  & 72.30\\
    \bottomrule
  \end{tabular}
\end{table}

\subsection{Class Prototype Analysis}
\begin{figure}[t]
    \centering
    \includegraphics[width=2.8in]{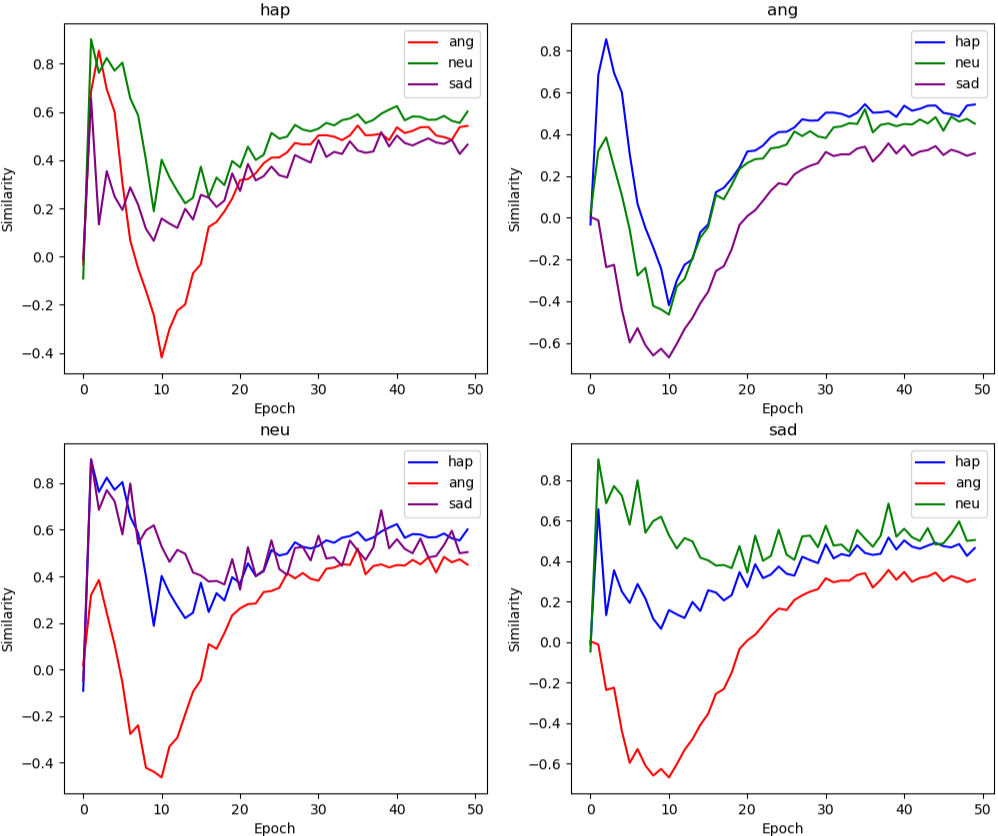}
    \caption{Similarity between class prototype embeddings during training.}
    \label{fig:proto}
\end{figure}

Figure~\ref{fig:proto} shows the trend of similarity between class prototype embeddings during the training process. In the initial 10 epochs of training, we observe a transient increase in the similarity between the neutral prototype and other class prototype vectors. This phenomenon arises from the localized occurrence of emotional expression within speech segments. Additionally, given the similarity in activation levels between happiness and neutrality, they exhibit the highest prototype similarity, aligning with prior finding \cite{liu22aa_interspeech}. Intuitively, anger and sadness represent contrasting emotions, thus exhibiting a declining trend in the similarity between their prototypes. In subsequent training stages, prototype-assigned pseudo labels are highly inaccurate, resulting in unlabeled data negatively impacting the model and thus reducing the quality of the learned class prototypes. Furthermore, all prototype embeddings are soft updated according to the proportion of each emotion in ambiguous soft labels, which leads to an increase in the similarity of the emotion prototypes. Ultimately, we observe that the class prototypes stabilize at a relatively steady state with a similarity greater than 0, affirming the inherent ambiguity in emotion expression.

\subsection{Pseudo Label Analysis}

\begin{figure}[t]
  \centering
  \includegraphics[width=2.8in]{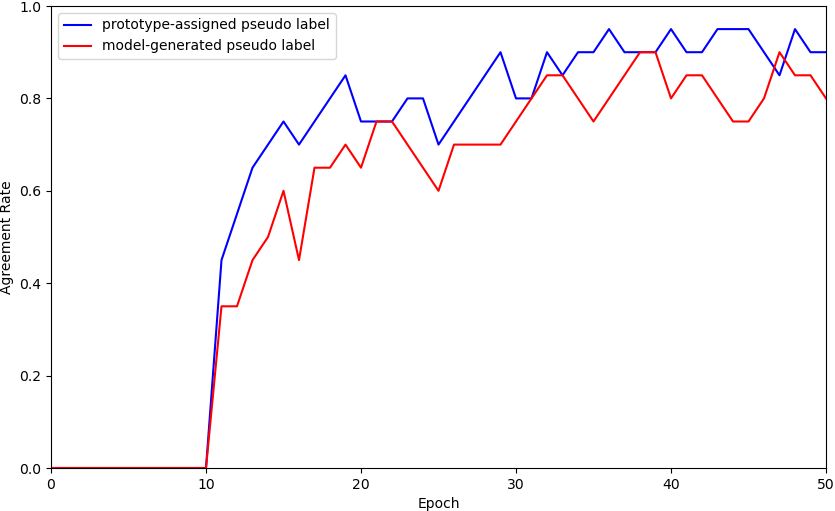}
  \caption{Blue (red) represents the agreement rate between model-generated (prototype-assigned) pseudo labels and ground-truth labels.}
  \label{fig:agree}
  \vspace{-10pt}
\end{figure}

Figure~\ref{fig:agree} shows agreement rate of model-generated pseudo labels and prototype-assigned pseudo labels with ground-truth labels. We observe that the upward trend in the agreement rate between prototype-assigned and ground-truth labels is significantly greater than the agreement rate between model-generated soft labels and ground-truth labels. Additionally, the prototype-assigned pseudo labels exhibit a higher agreement rate with the ground-truth labels. On the one hand, IPR could achieve ambiguous soft label annotation on a large amount of unlabeled data by training on a small amount of precise labeled data. On the other hand, it could assign more precise single labels while maintaining the desired recognition performance. 

\section{Conclusions}
In this paper, we propose a novel iterative prototype refinement framework leveraging class prototypes and contrastive learning for ambiguous SER. IPR capitalizes on the clustering effect induced by contrastive learning to generate optimal representations. These representations facilitate the refinement and updating of class prototype embeddings, thereby enhancing the precision of assigned pseudo labels. Consequently, improved pseudo labels contribute to the augmentation of representation quality, establishing a positive feedback loop. Through a comprehensive array of comparative experiments, ablation studies, and visualization analyses conducted on the IEMOCAP benchmark dataset, we substantiate the effectiveness of IPR.

\section{Acknowledgments}
This work was supported in part by NSF China (Grant No. 62271270).

\bibliographystyle{IEEEtran}
\bibliography{mybib}

\end{document}